\documentclass[12pt,preprint]{aastex}

\usepackage{graphicx}
\usepackage{layout}
\usepackage[latin1]{inputenc}
\usepackage{amssymb}
\usepackage{natbib}

\slugcomment{To appear in Ap. J.}
\shorttitle{High redshift quasar hosts}
\shortauthors{Kotilainen et al.}

\def\mincir{\ \raise -2.truept\hbox{\rlap{\hbox{$\sim$}}\raise5.truept %MC
\hbox{$<$}\ }}  %
\def\magcir{\ \raise -2.truept\hbox{\rlap{\hbox{$\sim$}}\raise5.truept %
\hbox{$>$}\ }}

\begin{document}

\title{
%Draft: {\bf \today}\\
The properties of quasar hosts at the peak of the quasar activity}

\author{Jari K. Kotilainen}
\affil{Tuorla Observatory, Department of Physics and Astronomy,
University of Turku, V\"ais\"al\"antie 20, FI--21500 Piikki\"o, Finland}
\email{jarkot@utu.fi}

\author{Renato Falomo}
\affil{INAF -- Osservatorio Astronomico di Padova, Vicolo dell'Osservatorio 5,
I-35122 Padova, Italy}
\email{renato.falomo@oapd.inaf.it}

\author{Roberto Decarli and Aldo Treves} 
\affil{Universit\`a dell'Insubria,
via Valleggio 11, I-22100 Como, Italy} \email{
roberto.decarli@mib.infn.it; aldo.treves@uninsubria.it}

\author{Michela Uslenghi}
\affil{INAF-IASF Milano, Via E. Bassini 15, Milano I-20133, Italy}
\email{uslenghi@iasf-milano.inaf.it}

\and

\author{Riccardo Scarpa}
\affil{Instituto de Astrofisica de Canarias, C/ Via Lactea, s/n E38205 - 
La Laguna (Tenerife), Spain}
\email{riccardo.scarpa@gtc.iac.es}

\begin{abstract}

We present near-infrared imaging obtained with ESO VLT/ISAAC of a sample of 
16 low luminosity radio-quiet quasars at the epoch around the peak of 
the quasar activity ($2 < z < 3$), aimed at investigating their host galaxies. 
For 11 quasars, we are able to detect the host galaxies and derive 
their properties, while for the other five quasars, upper limits to 
the host luminosity are estimated. The luminosities of the host galaxies of 
radio-quiet quasars at high redshift are in the range of those of massive 
inactive elliptical galaxies.

This work complements our previous systematic study of quasar hosts aimed 
to trace the cosmological luminosity evolution of the host galaxies up to 
$z \sim 2$ 
and extends our pilot study of a few luminous quasars at $z > 2$. 
The luminosity trend with cosmic epoch resembles that observed for 
massive inactive galaxies, suggesting a similar star formation history. 
In particular, both quasar host galaxies and massive inactive galaxies 
appear mostly assembled already at the peak age of the quasar activity. 
% It is found to be 
% consistent with the hypothesis of a passively evolving stellar population 
% generated in a burst at high redshift.
%$z_{\rm burst}\gtrsim3$. 
%, but is different from less 
%massive galaxies which have substantial mass growth by dry merging at 
%all epochs.
This result is of key importance for testing the models of joint formation 
and evolution of galaxies and their active nuclei.

\end{abstract}

\keywords{Galaxies:active --- Infrared:galaxies --- Quasars:general ---
galaxies: evolution}

\section{Introduction}

There is nowadays compelling evidence that the processes
of the formation and evolution of galaxies and the nuclear activity
are intimately linked. The most direct link at low redshift is the 
correlation of the mass of the central black holes with the
luminosity (mass) and the stellar velocity dispersion of the
spheroids where they reside, in both inactive and active galaxies
(see Ferrarese 2006 for a review). If this link holds also at higher redshift,
the observed population of high redshift quasars traces the existence of
$\sim 10^9$ M$_\odot$ supermassive BHs and massive spheroids at very
early ($<$ 1 Gyr) cosmic epochs \citep{fan03,willott03}. Moreover,
the strong cosmological evolution of the quasar population
\citep{dunlop90, warren94} is similar to the star formation history
in the Universe (Madau et al. 1998; see also Lapi et al., 2006).
According to the hierarchical merging scenarios for structure formation 
and evolution \citep[e.g.,][]{kauffmann00, dimatteo05}, 
the massive spheroidals should be the products of successive merger events, 
and the properties of quasar hosts (mass, luminosity, size) should show 
a redshift dependence. Moreover, the quasar activity is expected to
deposit large amounts of energy in the spheroid, possibly suppressing
star formation (e.g., Silk \& Rees 1998).

In this context, the direct detection and characterization of high
redshift quasar host galaxies is crucial to understand the joint
cosmic assembly and evolution of bulges of galaxies and their central black
holes. In particular, a key point is to probe the quasar host
properties up to and beyond the epoch of the peak of quasar activity
($2<z<3$).

Low redshift ($z \leq 0.5$) quasars are hosted in galaxies containing 
a luminous, massive bulge component that becomes dominant in radio-loud 
objects and at high nuclear luminosity 
\citep{hamilton02, dunlop03, pagani03, floyd04}. 
Their stellar populations are believed to be relatively old, especially 
in {\em very luminous} AGN compared to their inactive counterparts 
\citep{nolan01, dunlop03}. However,  there is some imaging and 
spectroscopic evidence for relatively young/intermediate age stellar 
populations in some low redshift AGN host galaxies, even in apparently 
quiescent ellipticals 
\citep{kauffmann03, jahnke04, kotilainen04, raimann05, hyvonen07b, letawe07, baldi08, hyvonen09}. 
%some of these samples are contaminated by late-type host galaxies, 
%REMOVED:
%This is particularly true in samples dominated by low-luminosity AGN 
%\citep[e.g. Seyferts and radio galaxies instead of quasars; c.f.][]{raimann05}.

The detection and characterization of the host galaxies of high
redshift quasars is challenging since the quasar luminosity
overwhelms the extended emission from the galaxy, especially in 
optical imaging, corresponding to rest-frame UV emission. Furthermore, 
the host galaxy surface brightness decreases rapidly with redshift. In
order to cope with these difficulties, imaging with high spatial
resolution and S/N together with a well defined point spread
function (PSF) are essential. Systematic reliable studies of the 
host galaxies of $z > 1$ quasars have thus became available only
recently thanks to the use of near-infrared (NIR) imaging,
where the nucleus-host luminosity ratio is more favourable,
allowing to detect the old stellar population at high redshift.
Concerning the stellar population, only few studies have been performed 
at high redshift (z $>$ 0.5) so far. 
%REMOVED:
%For example, Sanchez et al. (2004) 
%found a mixture of interacting blue galaxies, with colours typical of 
%star-forming galaxies, and normal, large, unperturbed ellipticals on the 
%red sequence. 

In the largest study of quasar host galaxies available up to now
\citep{falomo01, falomo04, kotilainen07}, we carried out systematic imaging
with the 8m Very Large Telescope (VLT) and the Infrared
Spectrometer And Array Camera (hereafter, ISAAC) of 32 quasars in
the redshift range $1 < z < 2$, to characterize their host galaxies.
The sample includes both radio-loud quasars (RLQs) and radio-quiet
quasars (RQQs), and it covers a wide range of the quasar
luminosities. We found that the luminosity evolution of both RLQ and
RQQ hosts is consistent with that of massive inactive ellipticals
undergoing passive evolution.

Beyond $z \sim 2$, detecting the host galaxies becomes extremely
difficult, even with the state-of-art observational techniques.
So far, only a few individual objects have been reported to be resolved
\citep{lehnert92, hutchings99, ridgway01, croom04, falomo05, peng06, falomo08, schramm08, villforth08}
mainly following three different approaches: {\bf (1)} observations from
space \citep[e.g.,][]{kukula01}, provide an excellent narrow PSF but
are usually limited by the small collecting area of the Hubble Space
Telescope (HST). 
Furthermore, the PSF of 
the images taken with the HST WFPC2 
are undersampled, leading to a systematic overestimate of the flux from 
the host galaxy (see Kim et al. 2008).  
Note that this undersampling does not apply to images taken with ACS 
and NICMOS. 
{\bf (2)} the extended emission from the host galaxies
is naturally magnified in gravitationally lensed quasars
\citep[e.g.,][]{peng06}. The drawback in this approach is that the
PSF of the lensed targets is difficult to characterize, and the lens
galaxy may contaminate the emission from the quasar host, making its
detection extremely uncertain. {\bf (3)} ground-based imaging with Adaptive
Optics (AO) \citep[]{croom04, falomo05, falomo08}
usually satisfies the severe constrains in the spatial resolution
required to disentangle the extended emission of the host galaxies
from the nuclear one. In our previous studies \citep{falomo05,
falomo08}, we obtained Ks-band images of quasars in the redshift
range $2 < z < 3$ using the AO system NACO at ESO VLT, which allowed
us to clearly resolve two RLQs and two RQQs. The drawback in this
approach is of statistical nature. In fact all current AO systems
require a very bright star to be present close to the target. This
therefore strongly limits the number of observable objects. This
limitation should be overcome with the next generation of AO systems
using artificial guide stars. {\bf (4)} Another possibility, that we shall 
address in this paper, is to focus on high-$z$ quasars of 
low luminosity where the nuclear-to-host luminosity ratio is 
more favorable. This ensures that, even without AO, the host galaxies 
can be detected and characterized from deep, NIR ground-based 
observations obtained in excellent natural seeing conditions. 
Moreover, the much larger field of view of non-AO images allows 
a significantly better characterization of the PSF from field stars.

In this paper, we present a deep, high resolution imaging study with
VLT/ISAAC of a sample of $2<z<3$ radio-quiet quasars with relatively 
{\it low nuclear luminosity}, aimed at the study of their host galaxies.
The structure of the paper is the following: In section 2 we describe
our sample. In section 3 we report the observations and data reduction
and in section 4 we describe the data analysis. In section 5 the
resulting luminosities of the quasar host are presented, together
with a comparison to literature data available in the same redshift range.
A discussion on the cosmic luminosity evolution of RQQ host galaxies
is presented in section 6.
%, and the relationship between host and nuclear luminosities.
Summary and directions for future work are given in section 7.
We adopt the concordance cosmology with
H$_0$ = 70 km s$^{-1}$ Mpc$^{-1}$, $\Omega_m$ = 0.3 and
$\Omega_\Lambda$ = 0.7.

\section{The sample}

\begin{figure}
\plotone{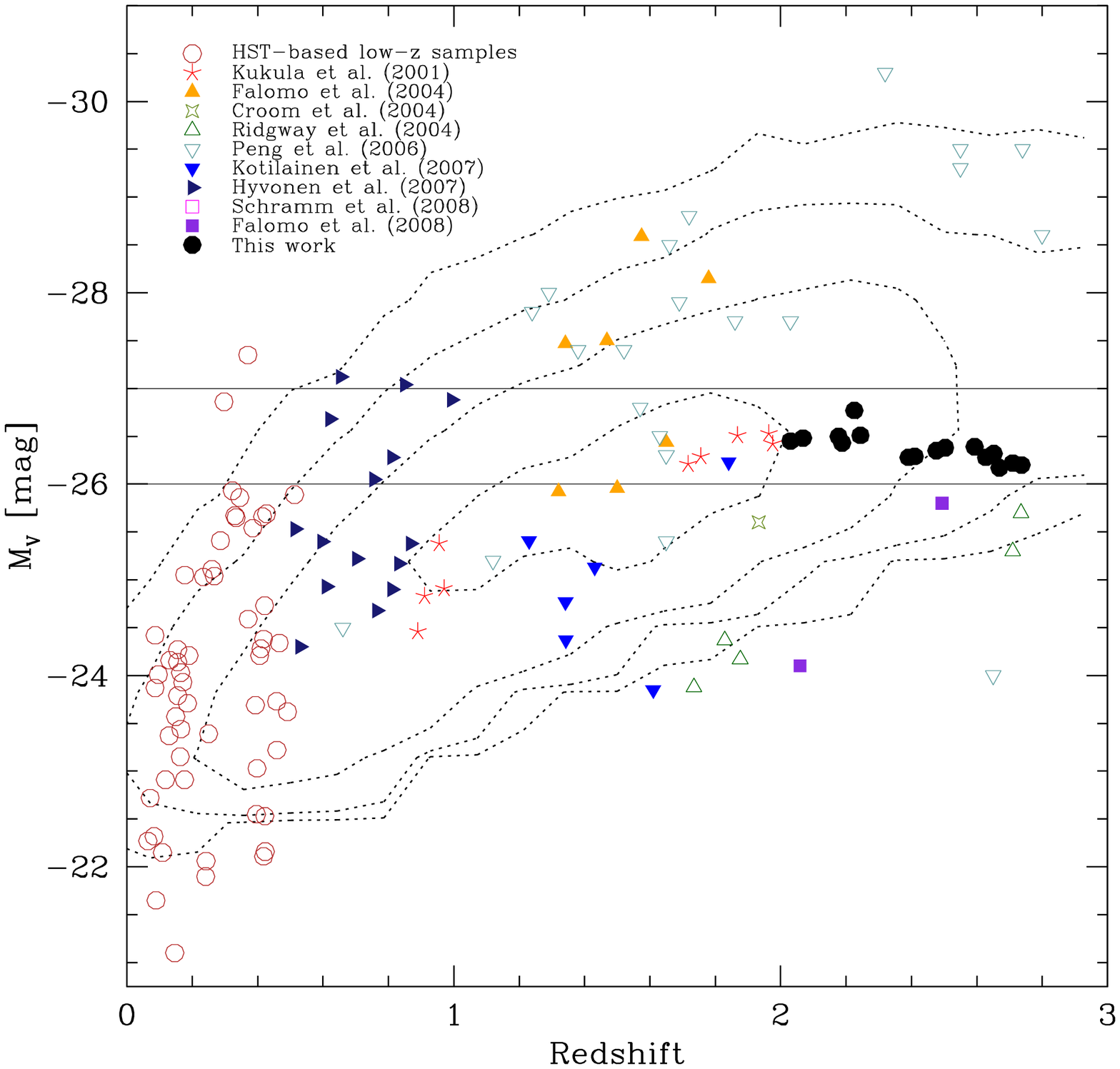}
\caption{
The rest-frame V-band absolute magnitude, as derived from the observed 
apparent V-band magnitude and assuming the Francis et al. (1991) template 
for $k$-correction, as a function of redshift for our sample objects 
(filled circles) and for similar objects in the literature (see the inserted 
symbol caption for details). Solid lines show the cut in absolute magnitude 
that we adopted in the target selection.
The contours (dotted lines) represent the distribution of all the quasars 
in the quasar catalogue of Veron-C\'etty \& Veron (2006), to which we 
refer to for details. The contour levels are logarithmically spaced, 
and reflect the number of quasars within each contour. Note that 
the absolute magnitudes of the targets from Peng et al. (2006) are corrected 
for the magnification due to gravitational lensing.
}\label{figveron}
\end{figure}

The sample of $2 < z < 3$ quasars was extracted from the AGN
catalogue of Veron-Cetty \& Veron (2006) requiring: {\bf (a)} {\em
rest-frame luminosity in the range}: $-26 > M_{\rm V} > -27$, $k$-corrected
assuming Francis et al. (1991) quasar SED template\footnote{Note that at $2 <
z < 3$ $k$-correction is $\leq 0.2$ mag for a typical
quasar SED.}, in order to maximize the likelihood for the host
galaxy detection; {\bf (b)} {\em at least two bright (R $<$ 16)
stars} within 40 arcsec of the quasar, and many other fainter stars
(R $<$ 20, comparable to the brightness of the quasars) within the
ISAAC $6.25$ arcmin$^2$ field of view, in order to provide a
reliable characterization of the PSF. With these constrains, the search 
produced $\sim 80$ quasars, from which we extracted a sample of 16 RQQs
with the most favourable observability
from Paranal (Chile), covering the full redshift range. There is no 
statistically
significant difference between the properties of the full and of the
observed samples. Fig. \ref{figveron} shows the distribution of
the observed quasars in the $z-M_{\rm V}$ plane, compared with the samples
of other similar studies available from literature and with the
envelope of all the quasars in the AGN catalogue of Veron-Cetty \&
Veron (2006).

\section{Observations}

Deep Ks-band images of the quasars were obtained using ISAAC
\citep{moorwood98}, mounted on UT1 (Antu) of VLT at the European
Southern Observatory (ESO) in Paranal, Chile. ISAAC is equipped with
a 1024 $\times$ 1024 pxl Hawaii Rockwell array, with a pixel scale
of 0\farcs148 px$^{-1}$, giving a field of view of $\sim 150 \times
150$ arcsec$^2$. The observations were performed in service mode under
photometric conditions in the period between 2006 March and
September. The journal of observations is given in Table
\ref{tabjournal}. The seeing, as derived from the FWHM size of stars
in each frame, was mostly excellent during the observations,
ranging from $\sim$0\farcs4 to $\sim$0\farcs6 (average and median
FWHM $\sim$0\farcs5, see Table \ref{tabjournal}). Note that at the 
redshift of the targets, observations in the Ks-band probe the host 
galaxy at rest-frame 5500 -- 7000 \AA, roughly the R-band, 
thus allowing an easy comparison with optical studies at lower redshift.

The images were secured using individual exposures of 2 minutes per
frame, and a jitter procedure \citep{cuby00}, which produces a set
of frames at randomly offset telescope positions within a box of 10
$\times$ 10 arcsec. The total
integration time was 38 minutes per target per observing block (OB).
All targets but one (2QZJ124029-0010) were observed in two or three
OBs of equal length. Since no significant discrepancy were found in
the sky surface brightness and the seeing from the PSF of field
stars between the different OBs for each target, we opted to combine
all the individual images of each target.

Data reduction was performed using our own improved version of the
ESO pipeline for jitter imaging data \citep{devillard01}. Each frame
was dark--subtracted and flat--fielded by a normalized flat field
obtained from twilight sky images. Sky subtraction was performed
using median averaged and scaled sky frames obtained combining
jittered exposures of the same field. Sky--subtracted images were
aligned to sub--pixel accuracy, and co--added. Combined images are
trimmed into a $850\times850$ pixel ($2.1\times2.1$ arcmin$^2$)
frame, covered by all the individual exposures. Finally, a
polynomial surface was fitted to the combined image, after masking
main sources, in order to remove spurious gradients in the
background counts due to sky variations during the integration.

Photometric calibration was performed from comparison with 2MASS 
magnitudes of bright stars available in the field.
The estimated internal photometric accuracy is $\sim0.1$ mag. 

\section{Two--dimensional data analysis}

Data analysis was carried out using AIDA (Astronomical Image
Decomposition and Analysis; Uslenghi \& Falomo, 2008), an IDL-based
software package designed to perform two--dimensional model fitting
of quasar images, providing simultaneous decomposition into nuclear
and host galaxy components. The applied procedure is described in
detail in Kotilainen et al. (2007), and briefly summarized here.

\subsection{PSF modeling}

The most critical part of the analysis is the determination of the
PSF model and the estimate of the background level around the target,
which may strongly affect the faint signal from the object. To model 
the PSF shape, we used field stars in each frame, selected on the 
basis of their FWHM, sharpness, roundness and S/N ratio, including 
bright, slightly saturated stars, in order to properly model the 
faint wing of the PSF. A careful check of the light profiles and 
contour plot of each star let us exclude marginally resolved galaxies  
and stars with close companions.

\begin{figure}
\plotone{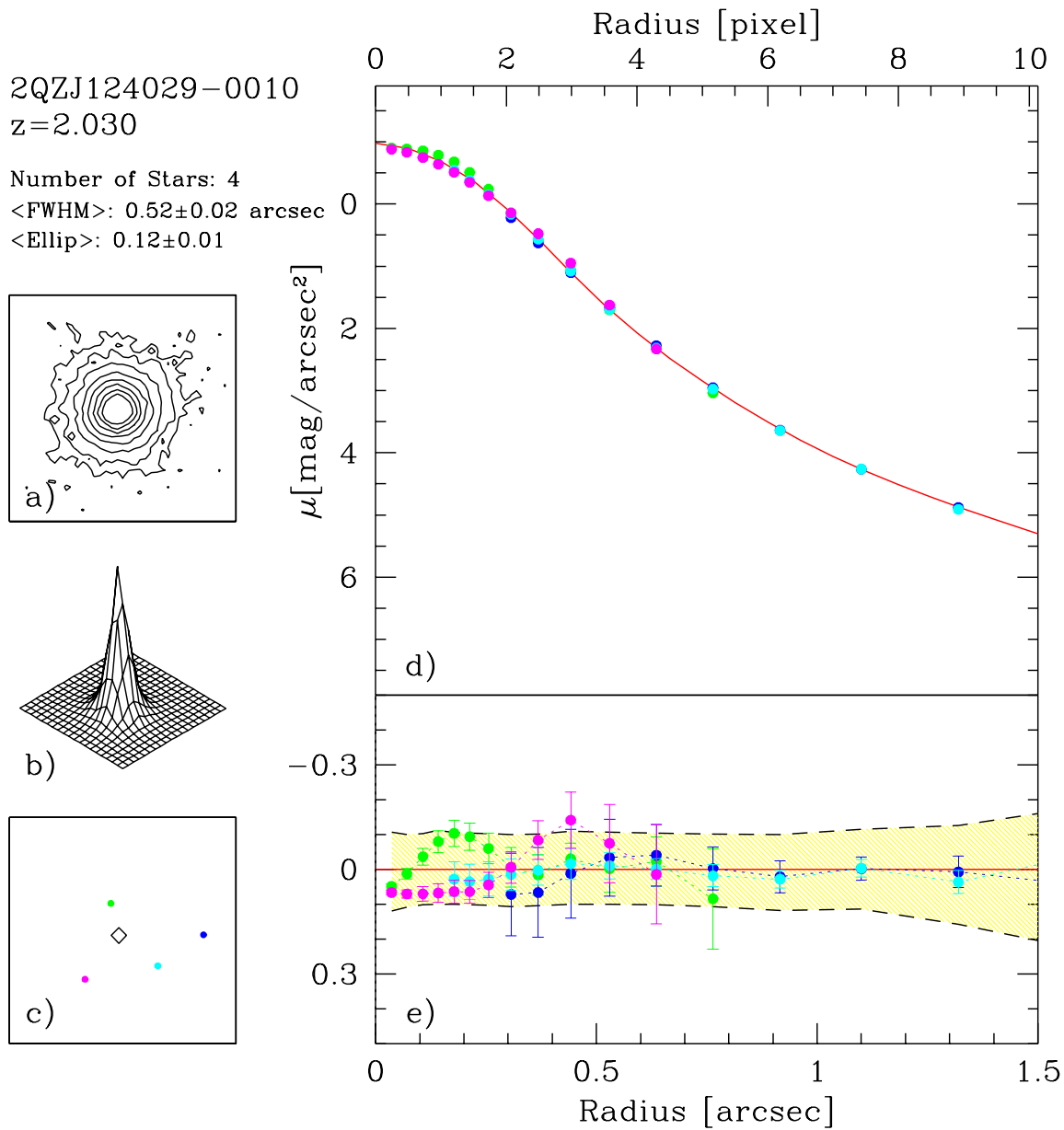}
\caption{
An example of the PSF modeling, for the frame of 2QZJ124029-0010.
The average and standard deviation of the FWHM and ellipticity
of the stars are listed. 
{\bf a)} the contour plot of a star in a $40\times40$ pixel 
($6\times6$ arcsec) box. 
{\bf b)} the same star in a 3D plot. 
{\bf c)} the positions of the stars (filled circles) used in the PSF modeling 
within the frame. The position of the target is also shown as 
an empty diamond. 
{\bf d)} the fitted PSF (solid line), compared to the observed star profiles 
(filled circles). 
{\bf e)} the residuals after the PSF modeling. The shaded area represents 
the estimated uncertainty in the PSF model, derived as described in the text.
}\label{figprofPSF}
\end{figure}

Each star was then modeled with four two-dimensional Gaussians,
representing the core of the PSF, and an exponential feature,
representing the extended wing of the PSF. Regions contaminated by nearby
sources, saturated pixels and other defects affecting the images
were masked out. The local background was computed in a circular
annulus centered on the star, and its uncertainty was estimated from
the standard deviation of the values computed in sectors of
concentric sub-annuli included in this area. Finally, the 
region used in the fit was selected by defining an internal and 
an external radius of a circular area, allowing the exclusion of the 
core of bright, saturated stars.
>From the comparison of the resulting light profiles (see Fig. 
\ref{figprofPSF}), no systematic dependence of the PSF was observed
in the field of view. Thus, the same model was fitted simultaneously
to all the usable stars of the image.

The uncertainty of the PSF model was estimated by comparing the
analytical fit with the individual observed star profiles, and 
adding a fixed term ($0.1$ mag/arcsec$^2$) to account for possible 
systematic effects due to underlying assumptions in the data reduction 
(e.g. Zero Point and PSF stability, and perfect alignments in 
the de-jitter procedure). Panel {\bf e)} in Fig. \ref{figprofPSF} 
shows the residual of PSF modeling together with the estimated uncertainty.

\subsection{Quasar host characterization}

\begin{figure}
\plotone{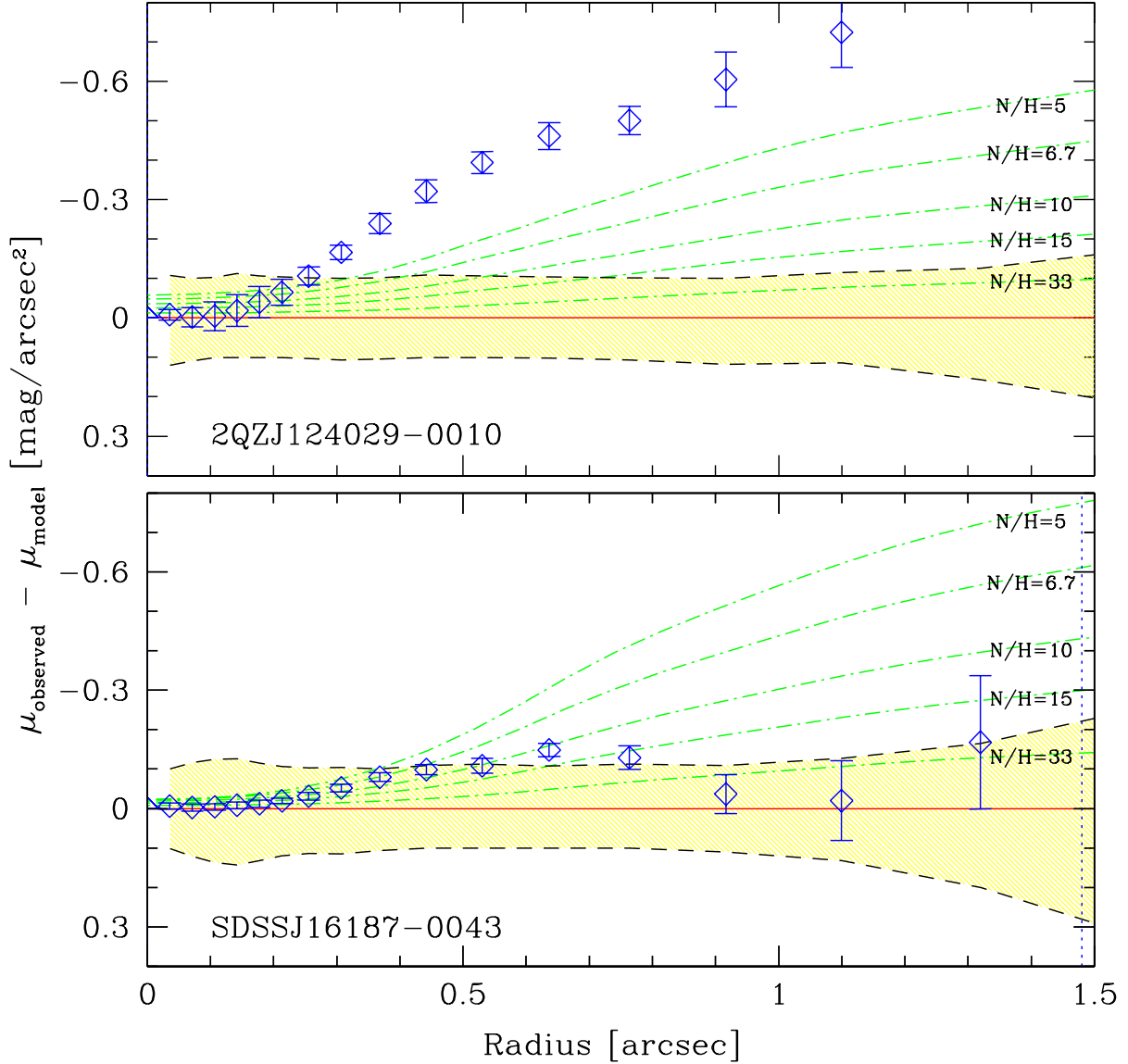}
\caption{
The residuals after the PSF fitting of the target (blue, empty diamonds) 
compared to the uncertainties in the PSF model (shaded area, 
see also panel {\bf e} in Fig. \ref{figprofPSF}). 
In the upper and lower panel, we show an example of a resolved target 
(2QZJ124029-0030) and an unresolved target (SDSSJ16187-0043), respectively. 
The green, dot-dashed lines refer to the simulated residual profiles for 
a quasar with a host galaxy following a de Vaucouleurs profile with 
$R_{\rm eff}=5$ kpc and various N/H ratios. 
}\label{figprofB}
\end{figure}

In order to evaluate whether the targets are resolved, we first fitted 
each quasar image using only the PSF model.
If the residuals revealed a significant excess over the PSF shape
in the $0.4$--$1.0$ arcsec range (where the contribution of 
the nuclear PSF quickly diminishes, while the contribution from 
the host galaxy signal becomes sufficiently high and detectable), 
the target was considered resolved. 
An example of the residuals after the PSF subtraction,
for a resolved target, is presented in Fig. \ref{figprofB}, upper panel.
The resolved targets were then fitted with a two component model (PSF+galaxy).
At high redshift, it becomes increasingly difficult to distinguish between
bulge and exponential disk models from the luminosity distributions.
We assumed that the host galaxies can be represented as elliptical 
galaxies with a de Vaucouleurs r$^{1/4}$ profile. This choice is 
motivated by the strong evidence at low redshift for the predominance of bulge
dominated hosts of quasars (e.g. Hamilton et al. 2002; Dunlop et al. 2003;
Pagani et al. 2003). If instead of a bulge-dominated model, we adopt 
an exponential disk law, all the objects can still be reasonably 
well fitted. The main difference is that the host galaxy luminosity become 
$\sim 0.3$ mag fainter for a disk model, but this does not significantly 
affect the results of this paper. 

In only two cases (2QZJ124029-0010 and 2QZJ222702-3205), the host galaxy 
detection was considered good enough to be able to firmly pin down 
the effective radius $R_{\rm eff}$. For all the other objects in the sample, 
we opted to set $R_{\rm eff}$ equal to 5 kpc in the modeling, following the 
indication of previous works \citep[e.g.,][]{falomo04}. 
We note that, within our adopted cosmological framework, 
5 kpc corresponds to $0.61$ arcsec at $z=2.5$, comparable with 
the typical angular resolution of our data.
In order to  evaluate the effect on the host luminosity of assuming a more 
compact host galaxy, we re-performed the modelling of all the resolved targets 
assuming $R_{\rm eff}$ = 2.5 kpc. With only two exceptions, the 
host galaxy magnitudes are consistent within 0.3 mag. 
No systematic offsets are found (the difference of average values = -0.03 mag).
The two exceptions are Q2125-4432 and 2QZ231751-3147 ($\Delta m=0.4$ and $0.8$
respectively), both of which have very faint host galaxies, 
close to the envelope of the unresolved targets, 
and therefore are likely to have larger uncertainties. 
Furthermore, the models of 2QZ231751-3147 have a highly 
elongated profile shape, suggesting that this object is peculiar.
The number of unresolved targets remains unchanged irrespective of the 
assumed $R_{\rm eff}$, since the resolved/unresolved classification only 
depends on the comparison between the observed light profile and the 
modelled PSF, independently of any other assumption on the shape of the host galaxy.

Targets showing no residuals in the light profile after the PSF subtraction 
exceeding the PSF model uncertainty were considered unresolved.
An example of an unresolved target is provided in Fig. \ref{figprofB}, 
lower panel. In these cases, we computed the profiles expected 
assuming a zero-ellipticity host galaxy with a de Vaucouleurs profile and 
$R_{\rm eff}=5$ kpc, and various N/H ratios. Among them, we adopted as 
the upper limit on the host galaxy luminosity the model that best matched 
the PSF model uncertainty.

\section{Results}

\begin{figure}
\begin{center}
\epsscale{0.8}
\plotone{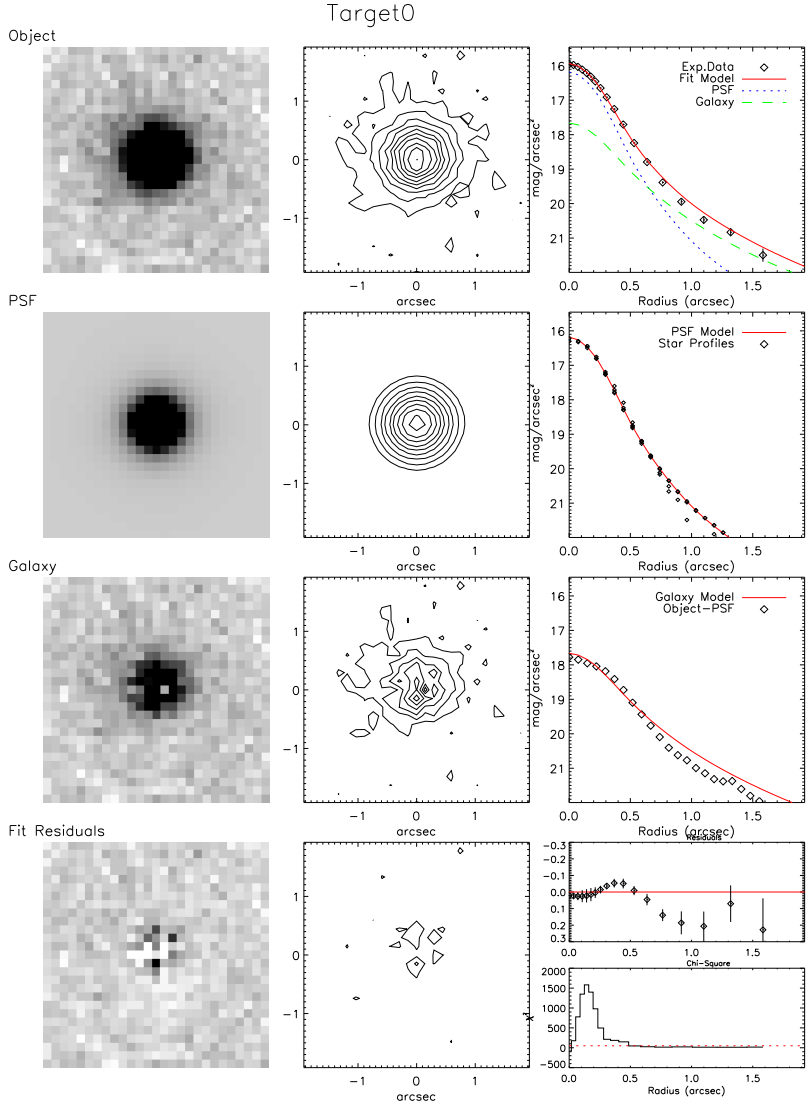}
\end{center}
\caption{
%Greyscale images (left), contour plots (middle) and radial luminosity 
%profiles (right) of t
The central 4 $\times$ 4 arcsec region surrounding 
the quasar 2QZJ124029-0010. In the left and middle panels, from top to bottom: 
(a) the original image, (b) the PSF model, (c) the host galaxy (PSF model 
subtracted from the observed profile), and (d) residuals of 
the fit. On the right, the top panel shows the observed 
radial profiles of the quasars (open diamonds), superimposed to the PSF model 
(blue dotted line) and an elliptical galaxy model convolved with its PSF 
(green dashed line). The red solid line shows the composite fit. 
The second panel shows the radial profiles of the stars compared to 
the PSF model (red solid line). The third panel shows the radial profile of 
the host galaxy after the PSF subtraction, while the bottom two panels show 
the residuals and the $\chi^2$ distribution of the fit. 
%Similar plots for all the targets in our sample 
%are available electronically at: 
%http://www.dfm.uninsubria.it/astro/qso_host/
}\label{figaida}
\end{figure}

In Fig. \ref{figaida}, we report the image of an example quasar, 
the best-fitting PSF, the residual after PSF subtraction, and
the residual after the fit with both the PSF and the galaxy model.
The fitted parameters are given in Table \ref{tabhost}. 
We have been able to resolve 11 out of the 16 quasars. 
In Table \ref{tabhost}, we also report the nuclear 
and the host galaxy absolute magnitudes and the nuclear to host (N/H) 
luminosity ratio for each quasar. All objects with N/H$\leq$12 are resolved, 
while no host galaxy is resolved in objects with N/H $>$ 12, 
highlighting its key role in the detection of the host galaxy emission.

To compare the properties of the quasar hosts at different redshifts,
it is preferable to compare data probing the same rest-frame wavelengths.
The Ks-band at redshifts $2 < z < 3$ corresponds to rest-frame
$\sim 5500-7000$ \AA{}, closely matching the R-band.
Therefore, in order to refer low and high redshift data to the same band
(and to minimize color and $k$-corrections), we transformed the observed
magnitudes into absolute magnitudes in the R-band. To perform the color and
$k$-correction transformations, we assumed an elliptical galaxy template
(Mannucci et al. 2001) for the host galaxy, and a composite quasar spectrum
(Francis et al. 1991) for the nucleus.

%Note that the $k$-correction in the observed Ks-band is virtually independent
%of the assumed host galaxy SED template in the redshift range
%considered here ($\Delta m <$ 0.1 mag; see the Appendix of
%Falomo et al. 2008), allowing a reliable measurement of
%the rest-frame luminosity.

All the resolved quasars in our study have host galaxies with luminosity 
ranging between M(R) $\sim-22.5$ and  M(R) $\sim-25.5$, 
i.e. corresponding to a range between $\sim M_*$ and $\sim M_*-3$, 
where $M_*$(R)$\sim-22.5$  is the R-band characteristic luminosity of 
the Schechter luminosity function for $2<z<2.5$ inactive galaxies 
(Marchesini et al. 2007), which is $\sim 1$ mag brighter than 
the one observed in the local Universe ($M_*$(R)$\sim-21.3$; 
see, e.g., Gardner et al. 1997; Nakamura et al. 2003). 
The average R-band host luminosity of
the resolved quasars is $<{\rm M(R)\, host}>=-24.2\pm0.8$.
Following the approach of statistical survival analysis proposed by
Feigelson \& Nelson (1985), we have computed the average M(R)host including
the upper limits of the host galaxy luminosities for the unresolved objects,
and obtain $<{\rm M(R)\, host}>=-23.8\pm0.5$. 
For a more thorough
discussion of this method, see Hyv\"onen et al. (2007a).
For reference, the average nuclear luminosity of the quasars is 
$<{\rm M(R)\, nucleus}>=-25.5\pm0.4$, and the average nucleus-to-host 
luminosity ratio $<{\rm N/H}>=6.7\pm1.2$.

\begin{figure}

\plotone{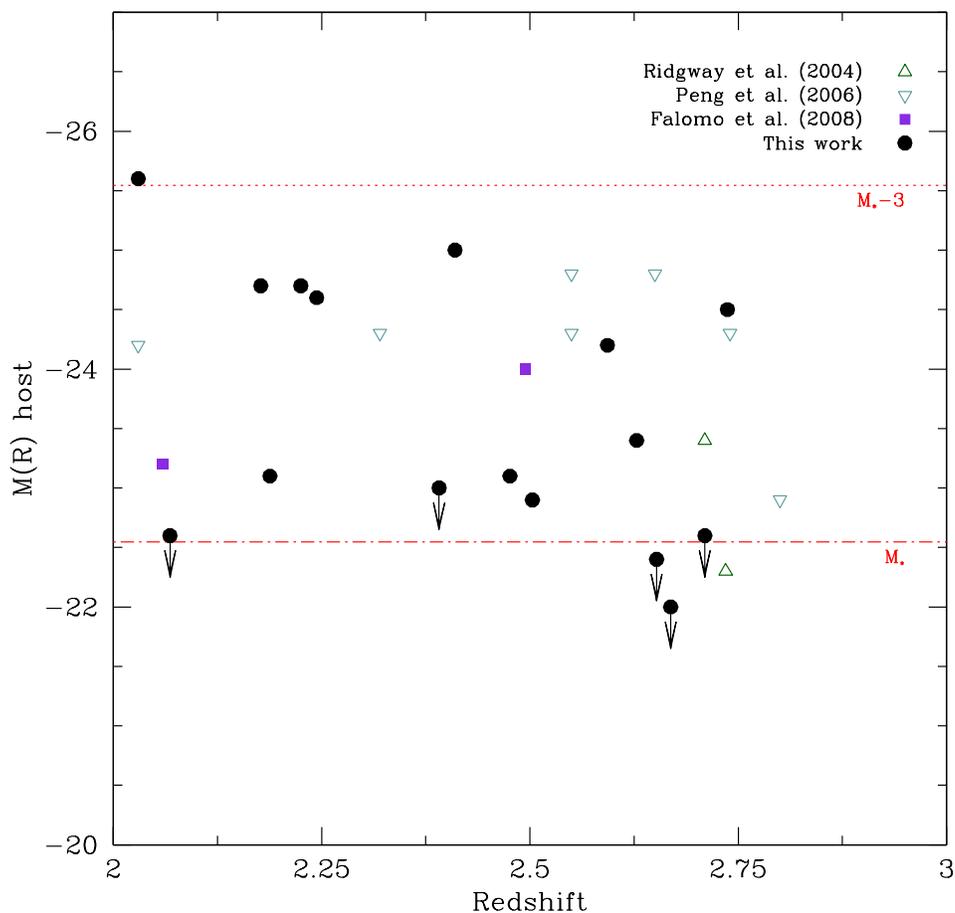}
\caption{
The host galaxy absolute magnitudes as a function of redshift for 
the targets in our work (filled circles) compared with
all the available literature data in the $2<z<3$ redshift range.
The arrows represent the upper limits of the host luminosity for 
the unresolved quasars in our study. 
The dash-dotted and dotted lines
show, for comparison, the values of $M_*$ and $M_*-3$, where $M_*$ the 
characteristic R-band magnitude of the luminosity
function of $2<z<2.5$ inactive galaxies (Marchesini et al. 2007).
}\label{figRMagH2z3}
\end{figure}

In Fig. \ref{figRMagH2z3} we compare the host galaxy luminosities derived 
in this paper with those available from the literature for quasars in 
the same redshift range. All the published magnitudes are converted to 
rest-frame R-band assuming the elliptical template from 
Mannucci et al. (2001), and adopting the same cosmological framework.
In the following, we compare our results with each available sample 
in the literature. 
The targets in Ridgway et al. (2001) were selected for having very low 
quasar luminosity. Their two $z\approx2.7$ quasars have 
M(R)host $= -22.3$ and $-23.4$ respectively, towards the faint end but 
consistent with the range of luminosities found in this paper.
The lensed quasars in Peng et al. (2006) have host galaxies with luminosities 
very similar to those found in our sample, with $<{\rm M(R)}>=-24.2\pm0.6$ 
(considering only the $2<z<3$ targets).
The same is true for the small number of resolved targets in the previous 
AO studies by our group (Falomo et al., 2005; 2008).
Villforth, Heidt \& Nilsson (2008) reported the marginal detection of 
the host galaxy of a single $z=2.75$ quasar. Its very low N/H ratio and 
spectral classification halfway between Type 1 and Type 2 AGN suggest that 
this object is an obscured quasar. Due to this peculiarity, we excluded it 
from further analysis.
Finally, Schramm et al. (2008) found very high host galaxy luminosities, 
M(R)host = $-25.8$, $-26.4$ and $-26.8$, for three quasars at $z=2.643$, 
$z=2.904$ and $z=2.933$, respectively. These luminosities are up to 
two magnitude brighter than in the other studies plotted in 
Fig. \ref{figRMagH2z3}. This discrepancy may be connected with the fact that 
these quasars are among the most luminous at this redshift range. 
Motivated by this peculiarity, we have re-analyzed the Ks-band data of 
Schramm et al., retrieved from the ESO Archive. Our re-analysis shows that 
all these quasars are unresolved. In particular, no signal from 
the host galaxy was found at radii larger than $\sim2$ arcsec (compare with 
Fig. 3 in Schramm et al., 2008). This is consistent with the presence of 
very bright nuclei in these quasars (about two magnitudes brighter than 
the brightest quasar in our sample). Furthermore, their observations were 
performed in only modest seeing conditions ($\sim$0.7 arcsec FWHM). 
Therefore, we have excluded the Schramm et al. data from 
Fig. \ref{figRMagH2z3} and the following analysis. 

\begin{figure}
\plotone{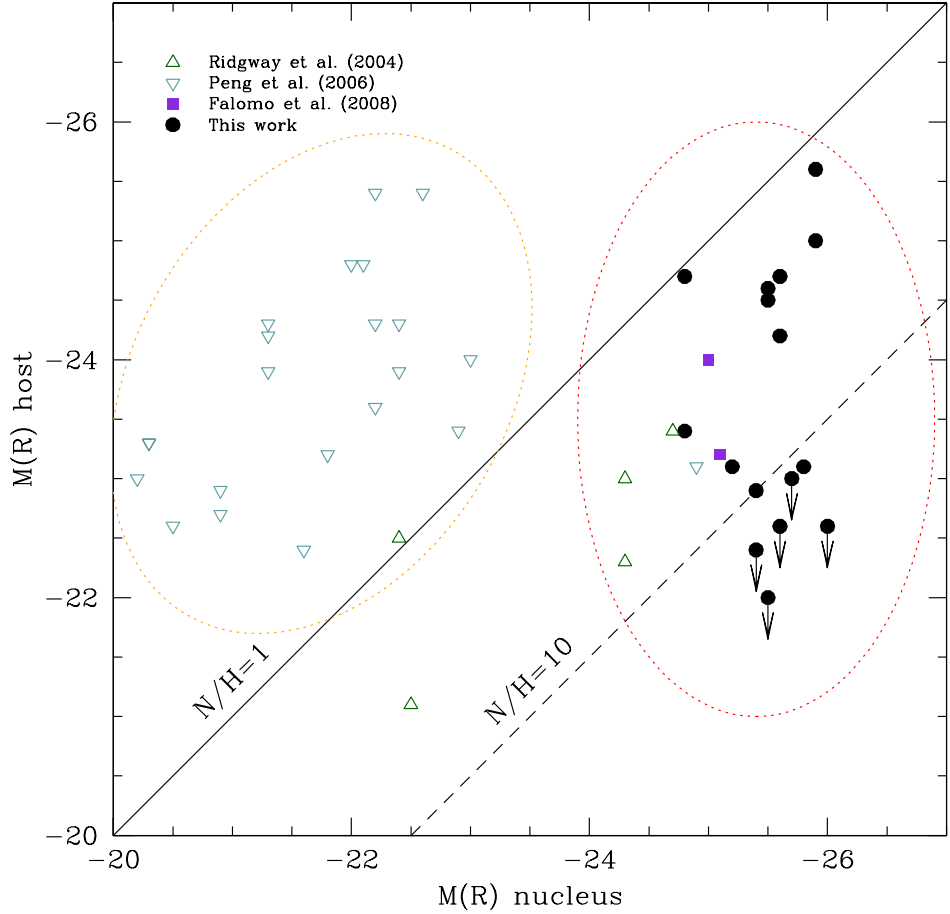}
\caption{
The absolute magnitude of the RQQ host galaxies compared to that of 
the nucleus in our sample (filled circles) and in other samples 
from the literature. The diagonal lines represent the loci of constant 
ratio between host and nuclear emission at N/H=1 (solid line) 
and N/H=10 (dashed line). For a similar range in M(R)host, 
tha data seem to fill two different regions of the plot, 
one with $-24>{\rm M(R) nucleus}>-26$ (data from Ridgway et al., 2001, 
Falomo et al., 2008 and this work) and one with $-20>{\rm M(R) nucleus}>-23$ 
(dominated by the sample of Peng et al., 2006). Note that having N/H$<$1, 
objects in the sample of Peng et al. should strictly be classified as 
lower luminosity AGNs (e.g. Seyferts) instead of quasars.
}\label{figN2Habs}
\end{figure}

As noted previously, the nucleus-to-host luminosity ratio of 
the resolved targets in our sample ranges between 1 and 12. 
If quasars emit with a narrow range in Eddington ratio and if 
the black hole mass is proportional to the luminosity of the host galaxy bulge 
(e.g., Magorrian et al., 1998), a correlation between the nuclear and 
the host galaxy magnitudes is to be expected. On the other hand, 
such a relation can fade due to several factors, both intrinsic 
(nuclear obscuration, beaming, and/or an intrinsic spread in 
the accretion rate) and extrinsic (e.g., lower level, non-quasar like 
nuclear activity may be relevant at low redshift; faint galaxies with 
very bright nuclei are difficult to resolve at high redshift), or can be 
an artifact of a narrowly sampled parameter space in the N/H ratio. 
Previous studies do not agree whether such a correlation exists 
(e.g., Dunlop et al., 2003; Pagani et al., 2003; Kotilainen et al., 2007).
Fig. \ref{figN2Habs} shows the comparison between the rest-frame $R$-band
host and nuclear absolute magnitudes for all the available RQQ host galaxies
in the $2<z<3$ range. 
No significant trend is observed between the host and the nuclear 
luminosity, neither considering only our sample nor comparing with all the 
available datasets, though we remark that the sampled parameter space 
in N/H strongly depends on the adopted observation technique, 
thus making a direct comparison between the various datasets difficult. 
We rather note that our new data, together with those published in 
Falomo et al. (2008) and most of the Ridgway et al. (2001) 
high redshift sample, fill a distinct region of the plot with respect to 
the data published in Peng et al. (2006). In other words, 
for a similar range in M(R)host, the quasars in Peng et al. have $\sim 3$ mag 
fainter nuclei, and their N/H$<1$, suggesting that the sample of Peng et al. 
is dominated by lower luminosity objects with non-quasar level, 
e.g. Seyfert type, nuclear activity. 

\section{Discussion}

\begin{figure}
%\plotone{RMag_z_RQQs.ps}
\plotone{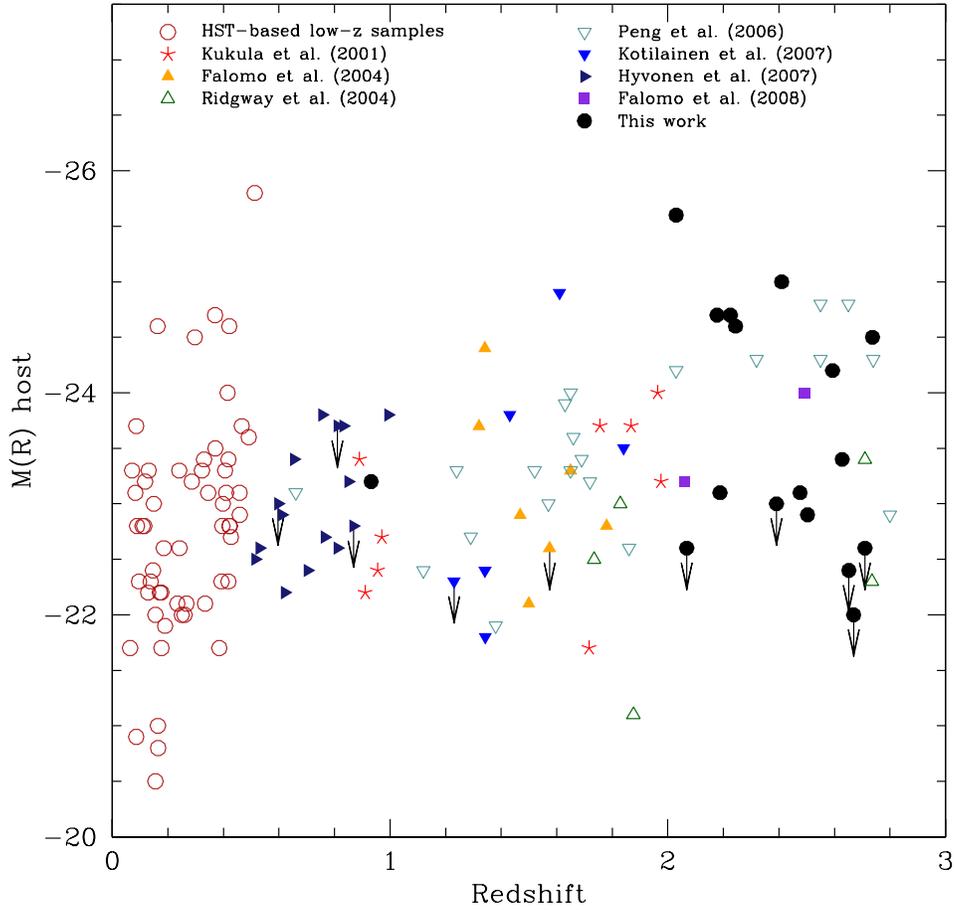}
\caption{
The redshift distribution of the host galaxy luminosity of 
radio-quiet quasars in this paper (filled circles) compared 
with all the available literature data. 
The arrows represent the upper limits of the host luminosity for 
the unresolved quasars. 
An overall $\sim1.5$ mag increase of the average M(R)host
is observed from $z=0$ to $z=3$. 
%Most of the data lie within the 
%curves expected for the passive evolution of a single-burst stellar 
%population with $z_{\rm burst}=5$ and normalized to $M_*$ and $M_*-2$
%at $z=0$ (dotted and dashed lines respectively).
}\label{figMevol}
\end{figure}

In Fig. \ref{figMevol} we extend the plot shown in Fig. \ref{figRMagH2z3}
to include all published RQQ host galaxy magnitudes at $0.5 < z < 2$
(Kukula et al. 2001; Ridgway et al. 2001; Falomo et al. 2004;
%Croom et al. 2004; 
Kotilainen et al. 2007; Hyv\"onen et al. 2007a),
and HST observations at $z<1$ (Bahcall et al., 1997; Hooper et al., 1997;
Boyce et al., 1998; Kukula et al., 2001;
Hamilton et al., 2002; Dunlop et al., 2003; Floyd et al., 2004; Labita et 
al., 2006). 
In order to treat these literature data homogeneously, we 
transformed the published apparent magnitudes to M(R)host following the 
procedure described above ($k$-correction, cosmology and color correction). 
Note that Croom et al. (2004) resolved the host galaxy of one quasar out of 
nine observed. Since they do not report upper limits on the magnitudes of 
the unresolved quasars, it is not possible to assess whether their 
unresolved hosts are consistent with the trend shown in Fig. \ref{figMevol}. 
Because of this, the contribution from the single detected host galaxy by 
Croom et al. is negligible in this context and was omitted. 

\begin{figure}
%\plotone{RMag_binz_bis.ps}
\plotone{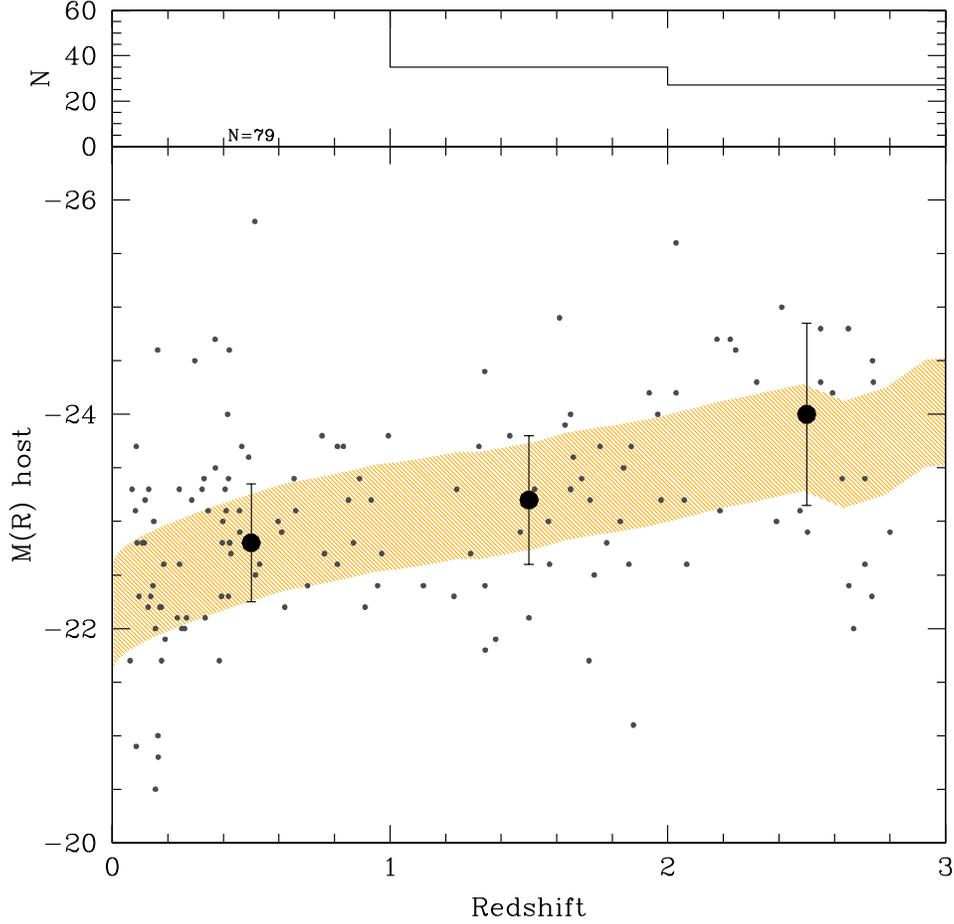}
\caption{
The redshift distribution of the host galaxy luminosity of 
radio-quiet quasars plotted in Fig. \ref{figMevol}, 
here binned as median averages at $z$ = 0.5, 1.5 and 2.5. Uncertainties
are the quartile values of the data distribution in each bin. The upper 
panel shows a histogram of the number of targets per bin.
The shaded area represents the evolution of the knee of inactive 
galaxy luminosity function from UKIDSS Ultra Deep Survey 
(Cirasuolo et al. 2007, 2008), $k$-corrected assuming the SED of a single-burst 
stellar population with $z_{\rm burst}=5$. The area extends from 
$M_*^{\rm UDS}$ to $M_*^{\rm UDS}-1$.
We note that the R--Ks color is poorly sensitive to the age of 
the stellar population: The difference between the adopted correction and 
a fixed $\sim 2.65$ color is $\lesssim0.15$ mag for $z$ ranging 
between 0 and 3.
}\label{figMevol2}
\end{figure}

We are now in a position to compare the observed general trend of
the luminosity of quasar host galaxies as a function of cosmic epoch with
the expectations of theoretical models of galaxy formation.
Fig. \ref{figMevol} shows that, despite the presence of a considerable 
scatter, a general trend is apparent, with the host galaxy luminosity of 
RQQs increasing by $\sim 1.5$ mag from the present epoch up to $z\sim3$. 
We stress that this trend is statistically significant. 
For example, considering data binned as median averages, in order to
be unaffected by the true (unknown) magnitude of the unresolved quasar hosts 
(see Figure \ref{figMevol2}), the linear best-fit 
of the relation is: M(R)host= -0.56$\pm$0.1 $\times$ z - 22.5$\pm$0.3, 
with $\chi^2$ = 1.3 (where the uncertainties are 3$\sigma$ errors from 
the $\chi^2$ maps 
and the $\chi^2$ is normalized to the number of degrees of freedom
). On the other hand, a fit with constant luminosity implies 
M(R)host = -23.2, with $\chi^2$ = 31.4.
%, while fitting with the normalization of an SSP track yields chi2 = 4.4. 
Therefore, the trend in Fig. \ref{figMevol} can not be reproduced with 
the host luminosity remaining constant with redshift. 

Most semi-analytical hierarchical models predict very few old, massive 
galaxies at high redshift (z $\geq$ 1) because in these models, large 
structures (massive galaxies) preferentially form at late epoch by 
continuous merging of smaller galaxies (e.g. Cole et al. 2000; 
Croton et al. 2006). 
This trend clearly disagrees with the observations presented here 
(Fig. \ref{figMevol}), and with the discovery of a substantial population 
of evolved, massive galaxies at z $>$ 1.5 (e.g. Daddi et al. 2005). 
Recent hierarchical models include feedback effects 
from supernovae and AGN to disentangle the baryon evolution from 
the hierarchical assembly of dark matter structures 
(e.g., Di Matteo et al., 2005; Bower et al. 2006; De Lucia et al., 2006; 
see Ellis, 2008 for a review on this topic). AGN feedback acts expelling 
gas from the galaxies and thus quenching star formation 
\citep{tremonti07, bundy08}. This effect is of fundamental
importance in massive galaxies, where AGN activity is preferentially observed 
(Kauffmann et al., 2003; Decarli et al., 2007; Gallo et al., 2008).

In order to probe whether inactive galaxies and quasar host galaxies 
have different star formation histories, we overplot in Fig. \ref{figMevol2} 
the trend observed for the characteristic luminosity 
$M_*^{\rm UDS}$ as a function of redshift, as derived from 
the UKIDSS Ultra Deep Survey (Lawrence et al., 2007).
The evolution of the luminosity function (Cirasuolo et al. 2007, 2008) is of 
the form:
\begin{equation}
\log M^{\rm UDS,AB}_*(z) = -22.26 - \left(\frac{z}{1.78\pm0.15}\right)^{(0.47\pm0.2)}
\end{equation}
We transformed the Ks${}^{\rm AB}$ magnitudes to the Ks-band assuming 
the correction used in the GOODS-ISAAC survey (Grazian et al. 2006): 
Ks${}^{\rm AB}$=Ks + $1.895$. Then we applied filter and $k$-correction as 
described above to compute the rest-frame R-band absolute magnitude. 
The result is shown in Fig. \ref{figMevol2}. On average, the available data 
on RQQ host galaxy luminosities lie between 
$M_*^{\rm UDS}$ and $M_*^{\rm UDS}-1$ \emph{at all redshifts}. 
This reinforces the leading assumption that we make that quasars are harboured 
in luminous, massive galaxies. Furthermore, the dependence of $M_*$ on $z$ 
for quiescent galaxies (Cirasuolo et al. 2007, 2008) remarkably closely 
matches the trend observed in the M(R)host -- $z$ distribution for 
quasar host galaxies, and is consistent with the findings of other studies on 
the luminosity function of galaxies at high redshift 
(e.g., Marchesini et al. 2007). 
This dependence supports the idea that quasar hosts and massive inactive galaxies share 
a similar star formation history.

There is increasing evidence for a mass dependence of the star formation 
history of galaxies (e.g., Gavazzi et al., 1996; 2002; Cimatti et al., 2004; 
Treu et al., 2005a; 2005b; Thomas et al., 2005), in the sense that 
the more massive elliptical galaxies formed their stars in relatively 
shorter bursts of intense star formation and at higher redshift 
(z $\sim 2.5-5$), compared to less massive galaxies. 
Recent merger events, if and when they occur, do not significantly affect
the content of the stellar population of massive galaxies.
On the other hand, the formation of lower luminosity galaxies is shifted 
toward lower redshifts, having significant star formation 
%and substantial mass growth by dry merging 
present at all epochs (see e.g. Renzini 2006, 
Scarlata et al. 2007, Vergani et al. 2008).

We stress that while the case of the evolution of a single burst population 
formed at high redshift provides an adequate explanation for our results, 
we can not rule out other models, accounting for the accretion of the galaxy 
mass due to mergers and different stellar history recipes. While a na\"ive 
model with galaxies experiencing substantial mass accretion is not 
consistent with the data, a more complex model in which 
episodic star formation counter-weights the mass accretion of galaxies is 
formally acceptable. Nevertheless, 
we prefer an interpretation in which such a coincidence is not necessary.
A more detailed modelling of the evolution and the star formation history 
of the host galaxies, 
that takes into account the role of the active nucleus 
is required. However, that is beyond the scope of this paper.  
Finally, we note that the assumption of 
a different SED of the stellar population does not affect significantly 
our results, since the rest-frame R--Ks color addressed in this paper is 
fairly insensitive to the age of the stellar population.

\section{Summary and conclusions}

We have presented homogeneous high resolution NIR images for a sample
of 16 low luminosity RQQs in the redshift range $2 < z < 3$,
to characterize the properties and the cosmological evolution of their
host galaxies, in conjunction with data at lower redshift.
The host galaxy was resolved in eleven quasars out of the 16,
while the remaining were unresolved.
The RQQs in our sample have $<{\rm M(R)\, host}>=-23.8\pm0.5$,
$<{\rm M(R)\, nucleus}>=-25.5\pm0.4$ and  $<{\rm N/H}>=6.7\pm1.2$. 

Comparing our new data with literature data at $2<z<3$ and at lower redshift, 
we found that the quasar host galaxies follow the trend
in luminosity of massive inactive ellipticals (with M(R) ranging between
$M_*$ and $M_*-3$ at $2<z<3$, or $M_*-1$ and $M_*-4$ at z=0) 
undergoing passive evolution.
%with z$_{\rm burst}\gtrsim3$. 
This represents a fundamental constraint 
for models of galaxy formation and evolution and throws new light on 
our understanding of the interplay between the black holes, nuclear activity 
and the host galaxy evolution.

% The most compelling consequence of this
% is that quasar host galaxies appear mostly fully assembled even up to
% and beyond the age of maximum quasar activity, and that the last major
% merger event in the host galaxies must have occurred at very high redshift.
% More recent merger events do not significantly alter the global luminosity of
% quasar host galaxies. The overall trend of host galaxy luminosity as
% a function of cosmic epoch is very similar to that observed for inactive 
% galaxies up to $z\sim3$. 

To definitely pin down the host galaxy assembly history,
a large observational effort in increasing the sample of resolved
host galaxies at $z\gtrsim2-3$ is required, in order to increase
the statistical significance of the present results and to sample
a wider parameter space, whether in the N/H ratio,
in the host galaxy luminosity, or in the level of quasar radio emission.
Further observations with very high S/N and a very narrow, extremely
reliable PSF, eventually thanks to laser guide star assisted AO in
NIR will play a crucial role.
Improved knowledge about the colours and environments of quasar host galaxies 
will also be needed to test the predictions of galaxy formation models. 
Finally, in a forthcoming paper (Decarli et al., in preparation), we will 
compare the luminosity of the host galaxies of high redshift quasars with 
their black hole masses, as derived from the spectroscopy of their 
broad emission lines. This will enable a direct test of the evolution of 
the black hole mass -- host galaxy relations up to $z=3$.

\acknowledgments

We are grateful to Marco Scodeggio and Ruben Salvaterra for their 
help in the comparison with inactive galaxies, 
and for the anonymous referee for constructive criticism. 
This work was 
supported by the Academy of Finland (projects 8107775 and 8121122) 
and by the Italian Ministry for University and Research (MIUR) under 
COFIN 2002/27145, ASI-IR 115, ASI-IR 35, and ASI-IR 73.

This publication makes use of data
products from the Two Micron All Sky Survey, which is a joint
project of the University of Massachusetts and the Infrared
Processing and Analysis Center/ California Institute of Technology,
funded by the National Aeronautics and Space Administration and the
National Science Foundation. This research has made use of the
NASA/IPAC Extragalactic Database {\em(NED)} which is operated by the
Jet Propulsion Laboratory, California Institute of Technology, under
contract with the National Aeronautics and Space Administration.

Facilities: \facility{VLT:Antu(ISAAC)}

% z-M(V) distribution

% PSF model

% PSF residuals

%\addtocounter{figure}{-1}%

% RQQ RMag host comparison

% RQQ host evolution

% Luminosity evolution of RQQs vs inactive galaxies

% comparison to models
%\begin{figure}
%\plotone{check_models.ps}
%\caption{
%The linear best-fit to the data in Fig. \ref{figMevol2} (the thick solid line) 
%together with its 1,2, and 3 sigma uncertainties (shaded areas), compared to 
%various models for galaxy formation: 
%a) passive evolution of stellar population formed at high redshift
%%the SSP track with $z_{\rm burst}$ = 5 
%(long-dashed line); b) a model in which M/L ratio is constant and 
%the mass of the galaxy decreases with redshift as log(M$_host$) $\sim$-0.3 z 
%(e.g., Volonteri, Haardt \& Madau 2003), due to merger events (dotted line); 
%and c) a model similar to b) (i.e. in which log(M$_host$) $\sim$-0.3 z) 
%but the stellar population is completely formed at high redshift and 
%passively evolves ever since (the "dry merger" scenario, short-dashed line).
%}\label{checkmodels}
%\end{figure}

%  TABLE 1
% Journal of observations
%
\begin{deluxetable}{l c c c c c c c}
\tablecolumns{9} \tablewidth{0pc} \tablecaption {TABLE 1}
\tablecaption  {Journal of the observations}
\tablehead{\colhead{Quasar} & \colhead{z} & \colhead{V\tablenotemark{a}} &
\colhead{Date (2006)} &
\colhead{Seeing\tablenotemark{b}} &
\colhead{N\tablenotemark{c}} &
\colhead{ZP\tablenotemark{d}} &
\colhead{sky mag\tablenotemark{e}}\\
  &  & [mag] &  & [arcsec] &  & [mag] & [mag/arcsec$^2$] }
\startdata \hline
2QZJ124029-0010 & 2.030 & 19.76 & 11-Jul         &0.52 $\pm$0.02& 2&23.69$\pm$0.04 &12.85 \\
2QZJ133136-0002 & 2.710 & 20.74 & 10-Jul         &0.52 $\pm$0.01& 4&24.09$\pm$0.06 &13.30 \\
2QZJ143220-0215 & 2.476 & 20.38 & 17-May         &0.42 $\pm$0.01& 5&24.14$\pm$0.08 &13.39 \\
2QZJ144022-0122 & 2.244 & 20.02 & 21-May, 10-Jul &0.46 $\pm$0.05& 1&24.02$\pm$0.01 &13.53 \\
SDSSJ16187-0043 & 2.068 & 19.77 & 13-Apr         &0.45 $\pm$0.00& 4&24.00$\pm$0.08 &13.48 \\
Q2125-4432      & 2.503 & 20.39 & 13-May         &0.56 $\pm$0.01& 3&24.07$\pm$0.02 &13.00 \\
Q 2126-1148A    & 2.188 & 20.00 & 13-May         &0.45 $\pm$0.01& 5&24.12$\pm$0.02 &13.26 \\
2QZJ215539-3026 & 2.593 & 20.44 & 17-May         &0.43 $\pm$0.02& 6&24.10$\pm$0.08 &13.04 \\
2QZJ221139-3132 & 2.391 & 20.40 & 17-May         &0.47 $\pm$0.02& 5&24.14$\pm$0.09 &13.22  \\
Q2225-403       & 2.410 & 20.20 & 18-May, 11-Jun &0.46 $\pm$0.05& 1&24.05$\pm$0.04 &13.34  \\
2QZJ222702-3205 & 2.177 & 20.13 & 12-Jun         &0.45 $\pm$0.00& 2&24.09$\pm$0.00 &13.42 \\
2QZJ223048-2954 & 2.652 & 20.56 & 12-Jun         &0.51 $\pm$0.02& 4&24.14$\pm$0.14 &13.44  \\
2QZJ225950-3206 & 2.225 & 19.72 & 17-Jul         &0.56 $\pm$0.01& 2&23.95$\pm$0.10 &13.24  \\
2QZJ231751-3147 & 2.628 & 20.58 & 18-Jul, 2-Aug  &0.56 $\pm$0.01& 6&24.10$\pm$0.03 &13.22  \\
2QZJ232755-3154 & 2.737 & 20.73 & 14-Jul, 17-Jul &0.40 $\pm$0.01& 4&23.99$\pm$0.04 &13.11  \\
2QZJ233451-2929 & 2.669 & 20.72 & 19-Jun, 3-Aug  &0.46 $\pm$0.02& 2&24.16$\pm$0.12 &13.46  \\
\hline
\enddata
\tablenotetext{a}{Quasar V-band apparent magnitudes from Veron-Cetty \& Veron (2006).}
\tablenotetext{b}{The average and rms FWHM in arcsec of all stars in the frame. 
When only 1 non-saturated star was present, from the comparison of the PSF with
resolved targets we conservatively estimated a $\sim 0.05$ arcsec uncertainty
in the seeing.}
\tablenotetext{c}{Number of stars used in the seeing estimate.}
\tablenotetext{d}{Zero Point from 2MASS magnitude of field stars.}
\tablenotetext{e}{Sky surface brightness.}\label{tabjournal}
\end{deluxetable}

%  TABLE 3
% Properties of Quasars  and Host galaxies

\begin{deluxetable}{l c c c c c c}
\tablecolumns{8}
\tablewidth{0pc}
\tablecaption{Properties of the quasar host galaxies.}
\tablehead{
\colhead{Quasar} &  \colhead{z}   &
%\colhead{$m({\rm Ks})_{\rm nuc}$\tablenotemark{a}} & \colhead{$m({\rm Ks})_{\rm host}$\tablenotemark{b}} &
%\colhead{$M({\rm Ks})_{\rm nuc}$\tablenotemark{b}} & \colhead{$M({\rm Ks})_{\rm host}$\tablenotemark{b}} &
%\colhead{$M({\rm R})_{\rm nuc}$\tablenotemark{c}} & \colhead{$M({\rm R})_{\rm host}$\tablenotemark{c}} & 
%\colhead{N/H\tablenotemark{d}}
\colhead{$m({\rm Ks})$\tablenotemark{a}} & \colhead{$m({\rm Ks})$\tablenotemark{b}} &
%\colhead{$M({\rm Ks})$\tablenotemark{b}} & \colhead{$M({\rm Ks})$\tablenotemark{b}} &
\colhead{$M({\rm R}) $\tablenotemark{c}} & \colhead{$M({\rm R}) $\tablenotemark{d}} & 
\colhead{N/H\tablenotemark{e}}\\
 & &
\colhead{nuc} & \colhead{host} &
%\colhead{nuc} & \colhead{host} &
\colhead{nuc} & \colhead{host} & 
 
%  &  &   &   &     & & &
}
\startdata								    
\hline									    
2QZJ124029-0010 &2.030& 17.2  &$ 17.4 $& -25.9 & $ $ -25.6 &  $ $ 1.3 \\
2QZJ133136-0002 &2.710& 18.4  &$>21.3 $& -25.6 & $>$ -22.6 &  $>$16.0 \\
2QZJ143220-0215 &2.476& 18.3  &$ 20.5 $& -25.2 & $ $ -23.1 &  $ $ 6.7 \\
2QZJ144022-0122 &2.244& 17.7  &$ 18.7 $& -25.5 & $ $ -24.6 &  $ $ 2.2 \\
SDSSJ16187-0043 &2.068& 17.0  &$>20.4 $& -26.0 & $>$ -22.6 &  $>$23.1 \\
Q2125-4432      &2.503& 18.2  &$ 20.7 $& -25.4 & $ $ -22.9 &  $ $ 9.5 \\
Q 2126-1148A    &2.188& 17.2  &$ 20.1 $& -25.8 & $ $ -23.1 &  $ $12.0 \\
2QZJ215539-3026 &2.593& 18.3  &$ 19.5 $& -25.6 & $ $ -24.2 &  $ $ 3.4 \\
2QZJ221139-3132 &2.391& 17.6  &$>20.5 $& -25.7 & $>$ -23.0 &  $>$12.2 \\
Q2225-403       &2.410& 17.4  &$ 18.5 $& -25.9 & $ $ -25.0 &  $ $ 2.4 \\
2QZJ222702-3205 &2.177& 18.2  &$ 18.5 $& -24.8 & $ $ -24.7 &  $ $ 1.2 \\
2QZJ223048-2954 &2.652& 18.6  &$>21.5 $& -25.4 & $>$ -22.4 &  $>$15.8 \\
2QZJ225950-3206 &2.225& 17.5  &$ 18.6 $& -25.6 & $ $ -24.7 &  $ $ 2.4 \\
2QZJ231751-3147 &2.628& 19.2  &$ 20.4 $& -24.8 & $ $ -23.4 &  $ $ 3.5 \\
2QZJ232755-3154 &2.737& 18.6  &$ 19.5 $& -25.5 & $ $ -24.5 &  $ $ 2.5 \\
2QZJ233451-2929 &2.669& 18.5  &$>21.9 $& -25.5 & $>$ -22.0 &  $>$27.3 \\
\enddata
\tablenotetext{a} {Apparent magnitudes of the nuclei in the observed Ks-band.}
\tablenotetext{b} {Apparent magnitudes of the host galaxies in the observed Ks-band.}
\tablenotetext{c} {Absolute magnitudes of the nuclei in the R-band, $k$-corrected 
assuming the Francis et al. (1991) template; no correction for galactic extinction 
is applied.}
\tablenotetext{d} {Absolute magnitudes of the host galaxy in the R-band,
$k$-corrected assuming the elliptical galaxy template by Mannucci et al. (2001); 
no correction for galactic extinction is applied.}
\tablenotetext{e} {The N/H ratio, referred to the absolute R magnitudes.}
\label{tabhost}
\end{deluxetable}


\begin{thebibliography}{}
\bibitem[\protect\citeauthoryear{Bahcall et al.}{1997}]{bahcall97} Bahcall J.N., Kirhakos S., Saxe D.H., Schneider D.P., 1997, ApJ, 479, 642
\bibitem[\protect\citeauthoryear{Baldi \& Capetti}{2008}]{baldi08} Baldi,R.D., Capetti,A., 2008, A\&A, 489, 989
%\bibitem[\protect\citeauthoryear{Bennert et al.}{2008}]{bennert08} Bennert N., Canalizo G., Jungwiert B., et al., 2008, ApJ, 677, 846
\bibitem[\protect\citeauthoryear{Bower et al.}{2006}]{bower06} Bower R.G., Benson A.J., Malbon R., et al., 2006, MNRAS 370, 645
\bibitem[\protect\citeauthoryear{Boyce et al.}{1998}]{boyce98} Boyce P.J., Disney M.J., Blades J.C., et al., 1998, MNRAS, 298, 121
%\bibitem[\protect\citeauthoryear{Bruzual \& Charlot}{2003}]{bruzual03} Bruzual G., Charlot S., 2003, MNRAS, 344, 1000
\bibitem[\protect\citeauthoryear{Bundy et al.}{2008}]{bundy08}Bundy, K., Georgakakis, A., Nandra, K., et al., 2008, ApJ 681, 931 
\bibitem[\protect\citeauthoryear{Cimatti et al.}{2004}]{cimatti04} Cimatti A., Daddi E., Renzini A., et al., 2004, Nature, 430, 184
\bibitem[\protect\citeauthoryear{Cirasuolo et al.}{2007}]{cirasuolo07} Cirasuolo M., McLure R.J., Dunlop J.S., et al., 2007, MNRAS, 380, 585
\bibitem[\protect\citeauthoryear{Cirasuolo et al.}{2008}]{cirasuolo08} Cirasuolo M., McLure R.J., Dunlop J.S., et al., 2008, MNRAS, submitted (arXiv:0804.3471)
\bibitem[\protect\citeauthoryear{Cole et al.}{2000}]{cole00} Cole,S., Lacey,C.G., Baugh,C.M., Frenk,C.S., 2000, MNRAS 319, 168
\bibitem[\protect\citeauthoryear{Croom et al.}{2004}]{croom04} Croom S.M., Schade D., Boyle B.J., et al., 2004, ApJ, 606, 126
\bibitem[\protect\citeauthoryear{Croton et al.}{2006}]{croton06} Croton, D.J., Springel, V., White, S.D.M., et al. 2006 MNRAS 367, 864
\bibitem[\protect\citeauthoryear{Cuby et al.}{2000}]{cuby00} Cuby J.G., Lidman C., Moutou C., Petr M., 2000, Proc. SPIE, 4008, 1036
\bibitem[\protect\citeauthoryear{Daddi et al.}{2005}]{daddi05} Daddi,E., Renzini,A., Pirzkal,N., et al., 2005 ApJ 626, 680
\bibitem[\protect\citeauthoryear{Decarli et al.}{2007}]{decarli07} Decarli R., Gavazzi G., Arosio I., et al., 2007, MNRAS, 381, 136
\bibitem[\protect\citeauthoryear{De Lucia et al.}{2006}]{delucia06} De Lucia G., Springel V., White S.D.M., Croton D., Kauffmann G., 2006, MNRAS, 366, 499
\bibitem[\protect\citeauthoryear{Devillard}{2001}]{devillard01} Devillard, N. 2001 in Astronomical Data Analysis Software and Systems X, ASP Conf. Ser., 238, 10, 525
\bibitem[\protect\citeauthoryear{Di Matteo et al.}{2005}]{dimatteo05} Di Matteo T., Springel V., Hernquist L., 2005, Nature, 433, 604
\bibitem[\protect\citeauthoryear{Dunlop \& Peacock}{1990}]{dunlop90}  Dunlop J. S., Peacock J. A., 1990, MNRAS, 247, 19
\bibitem[\protect\citeauthoryear{Dunlop et al.}{2003}]{dunlop03} Dunlop J.S., McLure R.J., Kukula M.J., et al., 2003, MNRAS, 340, 1095
\bibitem[\protect\citeauthoryear{Ellis}{2008}]{ellis08} Ellis R.S., 2008, in First Light in the Universe, Saas-Fee Advanced Courses, 36, p. 259-364
\bibitem[\protect\citeauthoryear{Falomo et al.}{2001}]{falomo01} Falomo R., Kotilainen J.K., Treves, A., 2001, ApJ, 547, 124
\bibitem[\protect\citeauthoryear{Falomo et al.}{2004}]{falomo04} Falomo R., Kotilainen J.K., Pagani C., Scarpa R., Treves, A., 2004, ApJ, 604, 495
\bibitem[\protect\citeauthoryear{Falomo et al.}{2005}]{falomo05} Falomo R., Kotilainen J.K., Scarpa R., Treves A., 2005, A\&A, 434, 469
\bibitem[\protect\citeauthoryear{Falomo et al.}{2008}]{falomo08} Falomo R., Treves A., Kotilainen J.K., Scarpa R., Uslenghi M., 2008, ApJ, 673, 694
\bibitem[\protect\citeauthoryear{Fan et al.}{2003}]{fan03} Fan X., Strauss M.A., Schneider D.P., et al., 2003, AJ, 125, 1649
\bibitem[\protect\citeauthoryear{Feigelson \& Nelson}{1985}]{feigelson85} Feigelson E.D., Nelson P.I., 1985, ApJ, 293, 192
\bibitem[\protect\citeauthoryear{Ferrarese}{2006}]{ferrarese06} Ferrarese,L., 2006, Joint Evolution of Black Holes and Galaxies, eds. M. Colpi et al. (Taylor \& Francis Group), p. 1
%\bibitem[\protect\citeauthoryear{Fioc \& Rocca-Volmerange}{1997}]{fioc97} Fioc, M., Rocca-Volmerange, B., 1997, A\&A 326, 950
\bibitem[\protect\citeauthoryear{Floyd et al.}{2004}]{floyd04} Floyd D.J.E., Kukula M.J., Dunlop J.S., et al., 2004, MNRAS, 355, 196
\bibitem[\protect\citeauthoryear{Francis et al.}{1991}]{francis91} Francis P.J., Hewett P.C., Foltz C.B., et al., 1991, ApJ, 373, 465
\bibitem[\protect\citeauthoryear{Gallo et al.}{2008}]{gallo08} Gallo E., Treu T., Jacob J., et al., 2008, ApJ, 680, 154
\bibitem[\protect\citeauthoryear{Gardner et al.}{1997}]{gardner97} Gardner J.P., Sharples R.M., Frenk C.S., Carrasco B.E., 1997, ApJ Letters, 480, 99
\bibitem[\protect\citeauthoryear{Gavazzi et al.}{1996}]{gavazzi96} Gavazzi G., Pierini D., Boselli A., 1996, A\&A, 312, 397
\bibitem[\protect\citeauthoryear{Gavazzi et al.}{2002}]{gavazzi02} Gavazzi G., Bonfanti C., Sanvito G., Boselli A., Scodeggio M., 2002, ApJ, 576, 135
\bibitem[\protect\citeauthoryear{Grazian et al.}{2006}]{grazian06} Grazian A., Fontana A., de Santis C., et al., 2006, A\&A, 449, 951
\bibitem[\protect\citeauthoryear{Hamilton et al.}{2002}]{hamilton02} Hamilton T.S., Casertano S., Turnshek D.A., 2002, ApJ 576, 61
%\bibitem[\protect\citeauthoryear{Hasinger}{2008}]{hasinger08} Hasinger, G. , 2008, A\&A, 490, 905
\bibitem[\protect\citeauthoryear{Hooper et al.}{1997}]{hooper97} Hooper, E.J., Impey C.D., Foltz C.B., 1997, ApJ, 480, L95
%\bibitem[\protect\citeauthoryear{Hopkins \& Hernquist}{2006}]{hopkins06} Hopkins,P.F., Hernquist,L., 2006, ApJS 166, 1
\bibitem[\protect\citeauthoryear{Hutchings et al.}{1999}]{hutchings99} Hutchings J.B., Crampton D., Morris S.L.,Durand D., Steinbring E., 1999 AJ 117, 1109
\bibitem[\protect\citeauthoryear{Hyv\"onen et al.}{2007a}]{hyvonen07a} Hyv\"onen T., Kotilainen, J.K.,  \"Orndahl, E. , Falomo, R., Uslenghi, M. 2007a A\&A 462 525
\bibitem[\protect\citeauthoryear{Hyv\"onen et al.}{2007b}]{hyvonen07b} Hyv\"onen T., Kotilainen, J.K., Falomo, R. \"Orndahl, E. Pursimo,T., 2007b, A\&A 476, 723
\bibitem[\protect\citeauthoryear{Hyv\"onen et al.}{2009}]{hyvonen09} Hyv\"onen T., Kotilainen, J.K., Reunanen, J., Falomo, R., 2009, A\&A 499, 417
\bibitem[\protect\citeauthoryear{Jahnke et al.}{2004}]{jahnke04} Jahnke K., Kuhlbrodt B., Wisotzki L., 2004, MNRAS, 352, 399
\bibitem[\protect\citeauthoryear{Kauffmann \& Haehnelt}{2000}]{kauffmann00} Kauffmann G., Haehnelt M., 2000, MNRAS, 311, 576
\bibitem[\protect\citeauthoryear{Kauffmann et al.}{2003}]{kauffmann03} Kauffmann G., Heckman T.M., Tremonti C., et al. 2003, MNRAS 346, 1055
\bibitem[\protect\citeauthoryear{Kim et al.}{2008}]{kim08} Kim,M., Ho,L.C., Peng,C.Y., Barth,A.J., Im,M.,  2008, ApJS, 179, 283
\bibitem[\protect\citeauthoryear{Kotilainen \& Falomo}{2004}]{kotilainen04} Kotilainen J.K., Falomo R.,2004, A\&A, 424, 107 
\bibitem[\protect\citeauthoryear{Kotilainen et al.}{2007}]{kotilainen07} Kotilainen J.K., Falomo R., Labita M., Treves A., \& Uslenghi M., 2007, ApJ 660 1039
\bibitem[\protect\citeauthoryear{Kukula et al.}{2001}]{kukula01} Kukula M.J., Dunlop J.S., McLure R.J., et al., 2001, MNRAS, 326, 1533
\bibitem[\protect\citeauthoryear{Labita et al.}{2006}]{labita06} Labita M., Treves A., Falomo R., Uslenghi M., 2006, MNRAS, 373, 551
\bibitem[\protect\citeauthoryear{Lapi et al.}{2006}]{lapi06} Lapi A., Shankar F., Mao J., et al., 2006, ApJ, 650, 42
\bibitem[\protect\citeauthoryear{Lawrence et al.}{2007}]{lawrence07} Lawrence A., Warren S.J., Almaini O., et al., 2007, MNRAS, 379, 1599
\bibitem[\protect\citeauthoryear{Le Borgne et al.}{2004}]{leborgne04} Le Borgne D., Rocca-Volmerange B., Prugniel P., et al., 2004, A\&A, 425, 881
\bibitem[\protect\citeauthoryear{Lehnert et al.}{1992}]{lehnert92} Lehnert M.D., Heckman T.M., Chambers K.C., Miley G.K., 1992, ApJ, 393, 68
%\bibitem[\protect\citeauthoryear{Leitherer et al.}{1999}]{leitherer99} Leitherer, C., Schaerer, D., Goldader, J.D., et al., 1999, ApJS 123, 3 
\bibitem[\protect\citeauthoryear{Letawe et al.}{2007}]{letawe07} Letawe, G., Magain, P., Courbin, F., et al.,  2007, MNRAS 378, 83
\bibitem[\protect\citeauthoryear{Madau et al.}{1998}]{madau98} Madau P., Pozzetti L., Dickinson M. 1998 ApJ, 498, 106
\bibitem[\protect\citeauthoryear{Magorrian et al.}{1998}]{mago98} Magorrian, J.,Tremaine, S., Richstone, D., et al., 1998, AJ, 115, 2285
\bibitem[\protect\citeauthoryear{Mannucci et al.}{2001}]{mannucci01}Mannucci, F., Basile, F., Poggianti, B.M., et al. 2001, MNRAS, 326, 745
%\bibitem[\protect\citeauthoryear{Maraston}{2005}]{maraston05} Maraston,C., 2005, MNRAS, 362, 799
\bibitem[\protect\citeauthoryear{Marchesini et al.}{2007}]{marchesini07} Marchesini D., van Dokkum P., Quadri R., et al., 2007, ApJ, 656, 42
\bibitem[\protect\citeauthoryear{Moorwood et al.}{1998}]{moorwood98} Moorwood A., Cuby,J.-G., Biereichel,P., et al. 1998, The Messenger 94, 7
\bibitem[\protect\citeauthoryear{Nakamura et al.}{2003}]{nakamura03} Nakamura O., Fukugita M., Yasuda N., et al., 2003, AJ 125, 1682
\bibitem[\protect\citeauthoryear{Nolan et al.}{2001}]{nolan01} Nolan,L., Dunlop,J., Kukula,M., Hughes,D., Boroson,T., 2001, MNRAS, 323, 308
\bibitem[\protect\citeauthoryear{Pagani et al.}{2003}]{pagani03} Pagani C., Falomo R., Treves A., 2003, ApJ 596, 830
\bibitem[\protect\citeauthoryear{Peng et al.}{2006}]{peng06} Peng,C.Y., Impey,C.D., Rix,H.-W., et al. 2006, ApJ, 649, 616
\bibitem[\protect\citeauthoryear{Raimann et al.}{2005}]{raimann05} Raimann, D., Storchi-Bergmann, T., Quintana, H., Hunstead, R., Wisotzki, L., 2005, MNRAS 364, 1239
\bibitem[\protect\citeauthoryear{Renzini}{2006}]{renzini06} Renzini,A., 2006, ARA\&A, 44, 141
\bibitem[\protect\citeauthoryear{Ridgway et al.}{2001}]{ridgway01} Ridgway S., Heckman T., Calzetti D., Lehnert M., 2001, ApJ, 550, 122
\bibitem[\protect\citeauthoryear{Sanchez et al.}{2004}]{sanchez04} Sanchez,S.F., Jahnke,K., Wisotzki,L., et al., 2004, ApJ, 614, 586 
\bibitem[\protect\citeauthoryear{Scarlata et al.}{2007}]{scarlata07} Scarlata,C., Carollo,C.M., Lilly,S.J., et al.,  2007, ApJS, 172, 494
\bibitem[\protect\citeauthoryear{Schramm et al.}{2008}]{schramm08} Schramm M., Wisotzki L., Jahnke K., 2008, A\&A, 478, 311
\bibitem[\protect\citeauthoryear{Silk \& Rees}{1998}]{silk98} Silk J., Rees M.J., 1998, A\&A Letters, 331, 1
\bibitem[\protect\citeauthoryear{Thomas et al.}{2005}]{thomas05} Thomas D., Maraston C., Bender R., Mendes de Oliveira C., 2005, ApJ, 621, 673
\bibitem[\protect\citeauthoryear{Tremonti et al.}{2007}]{tremonti07} Tremonti, C. A., Moustakas, J., Diamond-Stanic, A.M., 2007, ApJ 663, L77
\bibitem[\protect\citeauthoryear{Treu et al.}{2005a}]{treu05a} Treu T., Ellis R.S., Liao T.X., van Dokkum P.G., 2005a, ApJ Letters, 622, 5
\bibitem[\protect\citeauthoryear{Treu et al.}{2005b}]{treu05b} Treu T., Ellis R.S., Liao T.X., et al., 2005b, ApJ, 633, 174
\bibitem[\protect\citeauthoryear{Uslenghi \& Falomo}{2008}]{uslenghi08} Uslenghi M. \& Falomo, R. 2008, Proc. Erice, in press.
\bibitem[\protect\citeauthoryear{Vergani et al.}{2008}]{vergani08} Vergani,D., Scodeggio,M., Pozzetti,L., et al., 2008, A\&A, 487, 89
\bibitem[\protect\citeauthoryear{Veron-Cetty \& Veron}{2006}]{veron06} Veron-Cetty M.P., Veron P. 2006, yCat, 7248, 0
\bibitem[\protect\citeauthoryear{Villforth et al.}{2008}]{villforth08} Villforth C., Heidt J. \& Nilsson K., 2008, A\&A 488, 133
% \bibitem[\protect\citeauthoryear{Volonteri et al.}{2003}]{volonteri03} Volonteri,M., Haardt,F., Madau,P., 2003, ApJ 582, 559 
\bibitem[\protect\citeauthoryear{Warren et al.}{1994}]{warren94} Warren, S.J., Hewett, P.C., Osmer, P.S. 1994, ApJ, 421, 412
\bibitem[\protect\citeauthoryear{Willott et al.}{2003}]{willott03} Willott C.J., Rawlings S., Jarvis M.J. \& Blundell K. M., 2003, MNRAS, 339, 173
\end{thebibliography}
\end{document}